\documentclass[12pt,preprint]{aastex}

\usepackage{euscript,amssymb,amsmath}
\usepackage{graphicx}

\newcommand{\be}[1]{\begin{equation}\label{#1}}
\newcommand{\ee}{\end{equation}}
\newcommand{\ba}[1]{\begin{eqnarray}\label{#1}}
\newcommand{\ea}{\end{eqnarray}}
\newcommand{\rf}[1]{(\ref{#1})}
\newcommand{\nn}{\nonumber}

\shorttitle{A unifying picture of helical and azimuthal MRI}
\shortauthors{Kirillov, Stefani and Fukumoto}

\begin{document}

\title{A unifying picture of helical and azimuthal MRI, and
the universal significance of the Liu limit}

\author{Oleg N. Kirillov, Frank Stefani}
\affil{Helmholtz-Zentrum Dresden-Rossendorf,
P.O. Box 510119, D-01314 Dresden, Germany\\
o.kirillov@hzdr.de, f.stefani@hzdr.de}

\author{Yasuhide Fukumoto}
\affil{Faculty of Mathematics, Kyushu University,
744 Motooka, Nishi-ku, Fukuoka 819-0395, Japan,
yasuhide@imi.kyushu-u.ac.jp}

\begin{abstract}
The magnetorotational instability (MRI) plays a key
role for cosmic structure formation by triggering turbulence
in the rotating flows of accretion disks that
would be otherwise hydrodynamically
stable. In the limit of small magnetic Prandtl number,
the helical and the azimuthal version of MRI
are known to be governed by a quite different scaling
behaviour than the standard MRI with a vertical applied
magnetic field.  Using the short-wavelength approximation
for an incompressible, resistive, and viscous rotating fluid
we present a unified description
of helical and azimuthal MRI, and we identify the universal character
of the Liu limit $2(1-\sqrt 2)\sim -0.8284$
for the critical Rossby number. From this
universal behaviour we are also lead to the prediction
of higher azimuthal wavenumber for rather
small ratios of azimuthal to axial applied fields.
\end{abstract}


\section{Introduction}

The magnetorotational instability (MRI)
is widely accepted as the main source of
turbulence and outward angular momentum transport
that is needed for the matter in accretion disks to spiral
inwards onto the central proto-star or black hole,
\cite{BH91}.
While the early work on MRI was mainly concerned with ideal
magnetohydrodynamics,
the last years have seen an increasing interest in the influence
of viscosity and electrical resistivity  on the MRI, \cite{PH08}.
A particular role is thought to be played by the magnetic
Prandtl number (${\rm Pm}$) which measures the ratio of viscosity to
resistivity. For accretion disks around black holes (BH),
\cite{BH08}
had discussed the transition from large values of ${\rm Pm}$,
in the vicinity of the BH, to small values in
the outer part of the disk, with ${\rm Pm}$ reaching unity
for approximately 100 Schwarzschild radii
(depending on several parameters, among them the
mass of the BH). By invoking a thermal runaway
process at the unstable interface
between regions with ${\rm Pm}>1$ and ${\rm Pm}<1$, the authors
associated this boundary with
the existence of high and low X-ray states, see
\cite{RM06}.

In recent years, a vivid discussion (\cite {LL07,FPLH,KK,OML11})
was devoted to the
possible decline of the angular momentum transport rate
with decreasing ${\rm Pm}$, and to the
intricate roles that are played here by the magnetic Reynolds number,
the detailed boundary conditions, and the stratification of the
disk.

Besides  this relevance to the
outer part of accretion disks, to protoplanetary disks
(\cite{TS08}),
and possibly even to planetary cores
(\cite{PDB08}),
the limit of low ${\rm Pm}$
has acquired some additional interest in
connection with the recent liquid metal
experiments devoted to the study of MRI
(\cite{SL04,SGGRSSH,NO10}).
While for the standard version of MRI (SMRI), characterized
by only a vertical field
being applied, the low ${\rm Pm}$ limit is rather smooth and
unspectacular (\cite{PH08}), the addition of an azimuthal field
leads to dramatic effects as revealed for the first time
by \cite{HR05}.
The arising helical MRI (HMRI), as we call it now, was shown to work
also in the inductionless limit since it is
governed by the Reynolds and Hartmann number, quite
in contrast to SMRI which is governed by the magnetic Reynolds number
and the Lundquist number. However, as it was early shown by
\cite{LIU06}, the functioning of HMRI is
limited to comparably steep
rotation  profiles with Rossby numbers
${\rm Ro}<{\rm Ro}_{Liu}=2(1-\sqrt 2)\sim -0.8284$ (which we
henceforth will call the
``Liu limit''), and does therefore not extend to the astrophysically
important Kepler profiles characterized by ${\rm Ro}=-0.75$.
This essential
limitation of the HMRI, together with a variety of further
parameter dependencies,
was confirmed in the
PROMISE experiment by \cite{SGGRSSH,SGGRSH,PRL}.
The intricate, though continuous, transition
between SMRI and HMRI which involves a spectral
exceptional point at which the
inertial wave branch coalesces with the branch of the
slow magnetocoriolis wave, was
clarified only recently by \cite{KS10,KS11}.

Another surprise in the limit of low $\rm Pm$ was,
for the case of a purely or strongly
dominant azimuthal magnetic field,
the numerical
prediction of a non-axisymmetric
version of MRI, working apparently in a
similar parameter region as HMRI (\cite{HTR10}).
Although the occurrence of MRI case
under the influence of a
purely azimuthal magnetic field had been studied much earlier, see
\cite{HGB95,OP,TQ97}, the crucial effect arose again for
the particular combination of low $\rm Pm$ and slightly
steeper than Keplerian shear profiles.
It has to be noticed that this azimuthal MRI (AMRI),
as we call it now, works for azimuthal
magnetic fields that are current-free in the considered
fluid, quite in contrast to the Tayler instability
(\cite{TA73,SE12}) that
is a pinch-type instability in a
current-carrying conducting medium.

The aim of this paper is to better understand why the
scaling behaviour of HMRI and AMRI, and their restriction to rather
steep rotation profiles, is so similar. In
order to clarify this point we restrict our work
completely to the short wavelength approximation (\cite{FL03,HF2003}),
keeping in mind that
some of our conclusions will need further confirmation
in more realistic simulations.

\section{Short wavelength
analysis of viscous, resistive  MRI for arbitrary azimuthal
wavenumbers}

We start from the equations of incompressible, viscous and resistive magnetohydrodynamics,
comprising the Navier-Stokes equation for the velocity field $\bf u$ and
the induction equation for the magnetic field $\bf B$,
\begin{eqnarray}
\frac{\partial {\bf u}}{\partial t}+{\bf u} \cdot
\nabla {\bf u}&=&\frac{{\bf B}\cdot
\nabla{\bf B}}{\mu_0 \rho} -\frac{\nabla P}{\rho} +\nu \nabla^2 {\bf u}\\
\frac{\partial {\bf B}}{\partial t}&=&{\bf B} \cdot \nabla {\bf u}-{\bf u} \cdot \nabla {\bf B}
+\eta \nabla^2 {\bf B},
\end{eqnarray}
where $P=p+\frac{{\bf B}^2}{2\mu_0}$ is the total
pressure, $\rho=const$
the density, $\nu=const$ the kinematic
viscosity,
$\eta=(\mu_0 \sigma)^{-1}$ the magnetic diffusivity,
$\sigma$ the conductivity of the fluid,
and $\mu_0$ the magnetic permeability of free space.
Additionally, the mass continuity equation for incompressible flows
and the divergence-free condition for the magnetic induction are used:
\begin{equation}
\nabla \cdot {\bf u} = 0,\quad  \nabla \cdot {\bf B}=0.
\end{equation}

In the following we consider a rotational flow in the gap between
the radii
$R_1$ and $R_2>R_1$, with an imposed magnetic field sustained by
currents external to the fluid (hence we disregard any
version of the Tayler instability and its combination with MRI).
Introducing the cylindrical coordinates $(R, \phi, z)$ we consider
the stability of a magnetized Taylor-Couette (TC) flow, i.e.
a steady-state background flow with
the angular velocity profile $\Omega(R)$ in
a (generally helical) background magnetic field,
\begin{equation}
 {\bf u}_0(R)=R\,\Omega(R)\,{\bf
e}_{\phi},\quad p=p_0(R), \quad {\bf B}_0(R)=B_{\phi}^0(R){\bf
e}_{\phi}+B_z^0 {\bf e}_z,
\end{equation}
with the azimuthal field component
\begin{equation}
B_{\phi}^0(R)=\frac{\mu_0 I}{2 \pi R},
\end{equation}
supposed to be produced by an axial current $I$.

The angular velocity profile of the background TC flow is known
to have the form
\begin{equation}
 \Omega(R)=a+\frac{b}{R^2},
\end{equation}
where $a$ and $b$ are defined as
\begin{equation}
a=\frac{\mu_{\Omega}-\hat \eta^2}{1-\hat \eta^2}\Omega_1,\quad
b=\frac{1-\mu_{\Omega}}{1-\hat \eta^2}R_1^2\Omega_1
\end{equation}
with the definitions
\begin{equation}
\hat \eta=\frac{R_1}{R_2},\quad \mu_{\Omega}=\frac{\Omega_2}{\Omega_1}.
\end{equation}
Introducing, as a measure of the steepness of the rotation
profile, the Rossby number $({\rm Ro})$,
\begin{equation}
{\rm Ro} =\frac{R}{2 \Omega} \frac{\partial \Omega}{\partial R}
\end{equation}
we find
\begin{equation}
a=\Omega(1+{\rm Ro}),\quad b=-\Omega R^2 {\rm Ro} .
\end{equation}

To study flow and magnetic field perturbations on the background of the magnetized TC flow we linearize the Navier-Stokes and induction equation
in the vicinity of the stationary solution by assuming
${\bf u}={\bf u}_0+{\bf u}'$, $p=p_0+p'$,
and ${\bf B}={\bf B}_0+{\bf B}'$ and leaving only
terms of first order with respect to the primed quantities.

Then, by using a short-wavelength approximation (the
details of the derivation will be published elsewhere) in the frame of the geometrical optics approach (see e.g. \cite{LS87,DS92,FL03,HF2003,LZ04,MB09})
we end up with a system of 4 coupled equations for the perturbations of
arbitrary azimuthal dependency
which generalize the corresponding equations derived in
\cite{KS10}. From those 4
 coupled equations, we can deduce the
dispersion relation
\begin{equation}
p(\gamma):=\det(H-\gamma E)=0
\label{de1}
\end{equation}
generated by the matrix
\begin{eqnarray}
H=\left(
    \begin{array}{cccc}
      -im\Omega-\omega_{\nu} & 2\alpha^2\Omega & i  \frac{m \omega_{A_{\phi}}+{\omega_A}}{\sqrt{\rho \mu_0} } & -\frac{2\omega_{A_{\phi}}\alpha^2}{\sqrt{\rho \mu_0} } \\
      -2\Omega(1+ {\rm Ro}) & -im\Omega-\omega_{\nu} & 0 & i\frac{m \omega_{A_{\phi}}+{\omega_A}  }{\sqrt{\rho \mu_0}} \\
      i(m \omega_{A_{\phi}}+{\omega_A})\sqrt{\rho \mu_0}  & 0 & -im\Omega-\omega_{\eta} & 0 \\
      2{\omega_{A_{\phi}}}{\sqrt{\rho \mu_0} } & i(m \omega_{A_{\phi}}+{\omega_A})\sqrt{\rho \mu_0} & 2\Omega {\rm Ro} & -im\Omega-\omega_{\eta} \\
    \end{array}
  \right)
\label{mH}
\end{eqnarray}
where we have used the following definitions for the viscous,
resistive, and the two Alfv\'en frequencies corresponding
to the vertical and the azimuthal magnetic field:
\begin{eqnarray}
\omega_{\nu}&=&\nu |{\bf k}|^2 \\
\omega_{\eta}&=&\eta|{\bf k}|^2 \\
\omega_A^2&=&\frac{k_z^2 {B_z^0}^2}{\rho \mu_0}\\
\omega_{A_{\phi}}^2&=&\frac{(B_{\phi}^0)^2}{\rho\mu_0 R^2}.
\end{eqnarray}
Note that $|{\bf k}|^2=k^2_R+k^2_z$, and $\alpha=k_z/|{\bf k}|$, where $k_R$, $m$, and $k_z$ are the radial, azimuthal, and axial wavenumbers of the perturbation. In the absence of the magnetic field, the dispersion relation determined by the matrix $H$ reduces to that derived already by \cite{KGD1966} for the non-axisymmetric perturbations of the hydrodynamical Couette-Taylor flow. Choosing, additionally, $m=0$, we reproduce the result of \cite{EY95}. In the presence of the magnetic fields and $m=0$, we arrive at the dispersion relation derived by \cite{KS10}.

The dispersion relation (\ref{de1}) generated by the matrix \rf{mH} can be rewritten
completely in terms of dimensionless numbers,
i.e. Rossby number $({\rm Ro})$,
magnetic Prandtl number $({\rm Pm})$,
ratio of the two Alfv\'en frequencies $(\beta)$,
Hartmann number $({\rm Ha})$, Reynolds number $({\rm Re})$
and a re-scaled azimuthal wavenumber $n$:
\begin{eqnarray}
{\rm Pm}&=&\frac{\nu}{\eta}=\frac{\omega_{\nu}}{\omega_{\eta}}\\
\beta&=&\alpha \frac{\omega_{A_{\phi}}}{\omega_A}\\
{\rm Re}&=&\alpha\frac{\Omega}{\omega_{\nu}}\\
{{\rm Ha}}&=&\alpha\frac{B_z^0}{k\sqrt{\mu_0 \rho \nu \eta}}\\
n&=&\frac{m}{\alpha}.
\end{eqnarray}

After re-scaling the spectral parameter as $\gamma=\lambda \sqrt{\omega_{\nu}\omega_{\eta}}$
we end up with the complex polynomial dispersion relation
\begin{equation}
p(\lambda)=a_0\lambda^4+(a_1+ib_1)\lambda^3+(a_2+ib_2)\lambda^2+(a_3+ib_3)\lambda+a_4+ib_4=0
\label{poly}
\end{equation}
with the coefficients:
\begin{eqnarray}
a_0&=&1 \nonumber \\
a_1&=&2\left(\sqrt{\rm Pm}+\frac{1}{\sqrt{\rm Pm}}\right) \nonumber \\
b_1&=&4 n {\rm Re} \sqrt{\rm Pm},\nonumber \\
a_2&=&2(\beta^2{\rm Ha}^2-3{\rm Re}^2{\rm Pm})n^2+4\beta{\rm Ha}^2 n
   +2(1+(1+2\beta^2){\rm Ha}^2)
   +4{\rm Re}^2(1+{\rm Ro}){\rm Pm}+\frac{a_1^2}{4}\nonumber\\
b_2&=& 6n{\rm Re}(1+{\rm Pm})\nonumber\\
a_3&=& a_1(\beta^2{\rm Ha}^2-3{\rm Re}^2{\rm Pm})n^2+2a_1\beta {\rm Ha}^2 n
   +a_1(1+(1+2\beta^2){\rm Ha}^2)
   +8{\rm Re}^2(1+{\rm Ro})\sqrt{\rm Pm}\nonumber\\
b_3&=& 4n^3\sqrt{\rm Pm}{\rm Re}(\beta^2{\rm Ha}^2-{\rm Re}^2{\rm Pm})\nonumber\\
   &+&2n{\rm Re}(4{\rm Pm}^2{\rm Re}^2(1+{\rm Ro})+(1+{\rm Pm})^2
   +2{\rm Pm}(1+{\rm Ha}^2))/\sqrt{\rm Pm}\nonumber\\
   &-&8(1-n^2)\beta{\rm Ha}^2{\rm Re}\sqrt{\rm Pm}\nonumber\\
a_4&=&((\beta^2{\rm Ha}^2-{\rm Re}^2{\rm Pm})n^2+2{\rm Ha}^2\beta n+
{\rm Ha}^2+2{\rm Pm}{\rm Re}^2)^2\nonumber\\
&+&2(2{\rm Re}^2{\rm Pm}{\rm Ro}+1)(({\rm Ha}^2\beta^2-
{\rm Re}^2{\rm Pm})n^2+2{\rm Ha}^2\beta n+{\rm Ha}^2)
-(1+{\rm Pm})^2{\rm Re}^2n^2\nonumber \\
&+&4{\rm Re}^2(1+{\rm Ro})-({\rm Ha}^2+2{\rm Pm}{\rm Re}^2)^2
+{\rm Ha}^4+1+4\beta^2{\rm Ha}^2\nonumber\\
b_4&=& 2{\rm Re}(1+{\rm Pm})(\beta^2{\rm Ha}^2-
{\rm Re}^2{\rm Pm})n^3+4{\rm Re}{\rm Ha}^2\beta(1+{\rm Pm})n^2\nonumber\\
&+&2{\rm Re}(2(1+{\rm Ro})(2{\rm Re}^2{\rm Pm}-
\beta^2{\rm Ha}^2(1-{\rm Pm}))+(1+{\rm Ha}^2)(1+{\rm Pm}))n\nonumber\\
&-&4\beta{\rm Ha}^2{\rm Re}(2+(1-{\rm Pm}){\rm Ro}).
\end{eqnarray}

Note again that this complex algebraic equation of 4th order
is valid for perturbations of arbitrary azimuthal wavenumber in
magnetized incompressible, viscous, resistive rotating fluids
exposed to current free axial and azimuthal magnetic fields. When $n=0$ it reduces to the dispersion relation of HMRI
derived by \cite{KS10}.

\begin{figure}[htp]
\begin{center}
 \includegraphics[angle=0, width=0.75\textwidth]{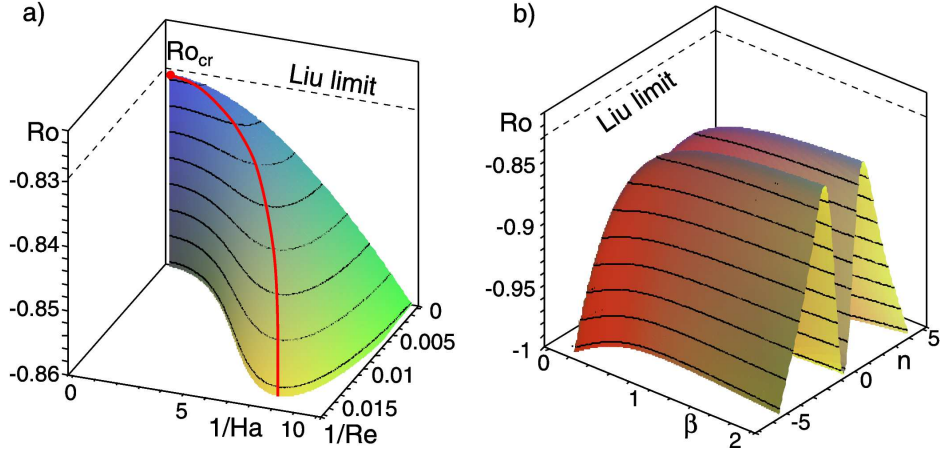}
\end{center}
\caption{Inductionless limit ${\rm Pm}\rightarrow 0$: (a) At the given $n=\sqrt{2}-1/20$ and $\beta=20$ the threshold of MRI in the $({\rm Re^{-1}},{\rm Ha^{-1}},{\rm Ro})$ space with the maximum at the singular (red) point $(0,0,{\rm Ro}_{cr}=2-2\sqrt{2})$. The red line leading to the maximum projects into a curve approximated in the vicinity of the origin by the scaling law \rf{sl}. (b) The threshold of MRI in the $(\beta,n,{\rm Ro})$ space with ${\rm Ha}=6$ and $\rm Re$ determined by the scaling law \rf{sl} reaches its upper bound at $\beta\rightarrow \infty$ (AMRI), which is still below the Liu limit ${\rm Ro}_{Liu}=2-2\sqrt{2}$. }
\label{fig0}
\end{figure}

\section{Inductionless limit}

Proceeding quite similar as in
\cite{KS10}, we apply the Bilharz criterion (\cite{Bilharz44}) to the complex polynomial \rf{poly} and derive the maximum Rossby number, at which flows are  prone to MRI, as a function of the remaining dimensionless numbers. In the following, we concentrate on the inductionless limit, i.e. we take the limit $\rm Pm \rightarrow 0$.

\begin{figure}[htp]
\begin{center}
 \includegraphics[angle=0, width=0.75\textwidth]{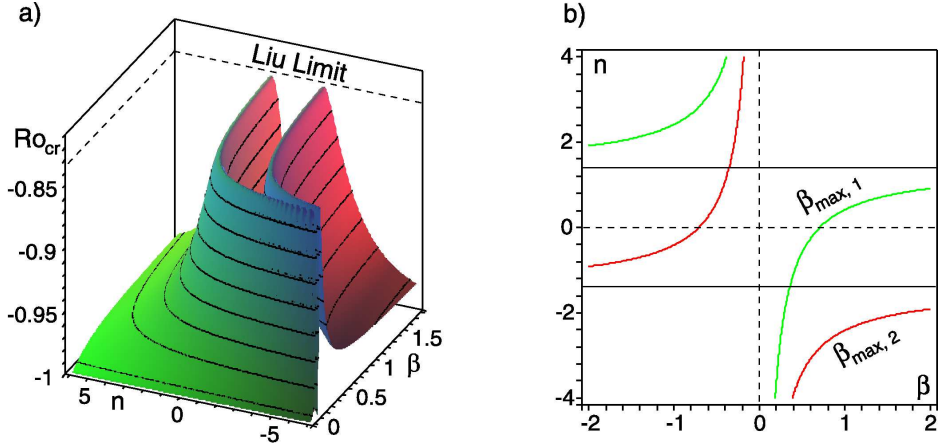}
\end{center}
\caption{(a) The critical Rossby number maximized with respect to $\rm Ha$ and $\rm Re$, given by Eq.~\rf{maineq}, in dependence on $\beta$ and $n$. (b) The lines (green) $\beta_{max,1}$ and (red) $\beta_{max,2}$ at which  the function ${\rm Ro}_{cr}(\beta,n)$ attains its maximal value ${\rm Ro}_{Liu}=2-2\sqrt{2}$.}
\label{fig1}
\end{figure}

After verifying the facts that the inductionless
threshold value of $\rm Ro$ increases monotonically with
${\rm Re}$ so that we can take the limit ${\rm Re}\rightarrow \infty$, and that in this limit the threshold value of $\rm Ro$ increases monotonically with ${\rm Ha}$, Fig.~\ref{fig0}(a), so that we can take the limit ${\rm Ha} \rightarrow \infty$, we obtain the following explicit expression for this maximized (with respect to $\rm Ha$ and $\rm Re$) critical Rossby number at the threshold of MRI in the inductionless limit:
\begin{eqnarray}
{\rm Ro}_{cr}(\beta,n)&=&\frac{4\beta^4+(\beta n+1)^4-(2\beta^2+(\beta n+1)^2)\sqrt{4\beta^4+(\beta n+1)^4}}
{2\beta^2(\beta n+1)^2}.
\label{maineq}
\end{eqnarray}
The maximum value of the Rossby number, ${\rm Ro}_{cr}$, is shown as a red dot in Fig.~\ref{fig0}(a).
When $n=0$, Eq.~\rf{maineq} reduces to the threshold of HMRI in the inductionless limit found in \cite{KS11}.
Given $n$ and $\beta$, the critical Rossby number calculated with finite values of $\rm Re$ and $\rm Ha$, is always below the majorating value, ${\rm Ro}_{cr}(\beta,n)$, determined by Eq.~\rf{maineq}, see Fig.~\ref{fig0}(b). With the increase of {\rm Ha} and $\rm Re$ constrained by the scaling law
\be{sl}
{\rm Re}=2(1+\sqrt{2})\beta^3{\rm Ha}^3,
\ee
the threshold shown in Fig.~\ref{fig0}(b) tends to the majorating surface ${\rm Ro}_{cr}(\beta,n)$ shown in Fig.~\ref{fig1}(a).

The remarkably simple dependence \rf{maineq} of ${\rm Ro}_{cr}$,
only on the ratio $\beta$ of
azimuthal to axial field and on the rescaled azimuthal wavenumber $n$,
relies on the appropriate choice of the dimensionless parameters,
in particular on ``hiding'' the wavenumber ratio $\alpha$ in them.
Assume we have fixed the sign of $\beta$, the  functional
dependence of the maximum critical
Rossby number \rf{maineq} on $\beta$ and $n$ has a two-saddle line structure
as shown in Fig.~\ref{fig1}(a).

Now, the most important result of this paper is that
both saddle lines have  the same  height everywhere, namely the Liu limit
${\rm Ro}_{Liu}=2(1-\sqrt{2})$, independent on the particular
combination of $\beta$ and $n$.
From Eq.~(\ref{maineq}) it can be proved that these saddle lines, with ${\rm
Ro}_{Liu}=2-2\sqrt{2}$,
are governed by the two equations
\ba{maxima}
\beta_{max,1}(n)&=&\frac{1}{\sqrt{2}-n},\quad{\rm Ro}_{cr}(\beta_{max,1})=2-2\sqrt{2}\nn\\
\beta_{max,2}(n)&=&\frac{-1}{\sqrt{2}+n},\quad{\rm Ro}_{cr}(\beta_{max,2})=2-2\sqrt{2}.
\ea
In Fig.~\ref{fig1}(b) the curves $\beta_{max,1}(n)$ and $\beta_{max,2}(n)$  are shown in green and red colors, respectively. Note that according to the first of Eqs.~\rf{maxima}, $n=0$ (HMRI) corresponds to $\beta=1/\sqrt{2}$ which being substituted into Eq.~\rf{sl}, yields the following scaling law for the optimum combination of $\rm Re$ and $\rm Ha$ in HMRI (\cite{KS10})
\be{sl2}
{\rm Re}=\frac{2+\sqrt{2}}{2}{\rm Ha}^3.
\ee

Therefore, even in the case of non-axisymmetric perturbations, the maximum possible value of the Rossby number prone to the magnetorotational instability caused by the helical magnetic field in the inductionless limit is still ${\rm Ro}_{Liu}=2-2\sqrt{2}$, exactly as in the case of HMRI which is an instability with respect to the axisymmetric perturbations $(n=0)$. The relations \rf{maxima} between $\beta$ and $n$ that correspond to the Liu limit give a sort of the resonance conditions between the components of the wavevector of the three-dimensional perturbation and the components of the helical magnetic field.

\begin{figure}[htp]
\begin{center}
\includegraphics[angle=0, width=0.55\textwidth]{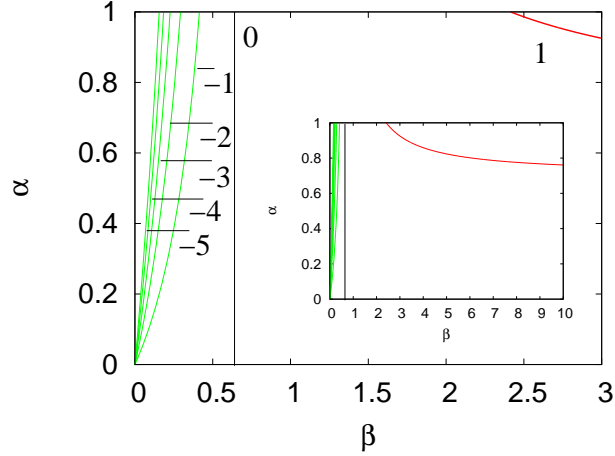}
\end{center}
\caption{Range of $\alpha$ for the different $m$ modes, in dependence on $\beta$.}
\label{fig2}
\end{figure}

On the basis of the equations \rf{maxima} connecting $\beta$ and the
rescaled azimuthal wavenumber
$n$, we can ask now for the structure of the solution in terms
of the original, ``physical'' azimuthal wavenumber $m$.
From the definition $\alpha=k_z/|{\bf k}|$
we see immediately that $\alpha$ can take on only values between
-1 and +1. The solution structure can thus be visualized as in Fig.~\ref{fig2}.
For example, from the first of the Eqs.~\rf{maxima} we see that
for large values of $\beta$ (AMRI), the only possible
integer solution is the $m=1$ mode,
whose corresponding wavenumber ratio converges than to $\alpha=1/\sqrt{2}$, see also Fig.~\ref{fig1}(b).
There is a lower limit of $\beta=1+\sqrt{2}\sim 2.41$ for this $m=1$ mode.
Lowering $\beta$ further, we find next the $m=0$ mode (HMRI) to dominate at the Liu limit,
restricted only to $\beta=1/\sqrt{2}$, \cite{KS10}.
Interestingly, decreasing $\beta>0$ even further to zero we find a sequence
of higher azimuthal modes with the sign of $m$ that is opposite to the sign of $\beta$, which indicates a kind of resonance phenomenon.

Since $\alpha$ enters also the definition of $\beta$ it might be instructive to
illustrate the mode structure also in dependence on
$\beta/\alpha=\omega_{A_{\phi}}/\omega_A$.
From Eqs.~\rf{maxima} we derive
\be{ab}
\alpha=\pm\frac{\sqrt{2}}{2}\left(m+\frac{\omega_A}{\omega_{A_{\phi}}}\right),
\ee
where the positive sign corresponds to the first of Eqs.~\rf{maxima} and the negative sign to the second one.
This means that with a given azimuthal wavenumber $m$, two axial wavenumbers, $k_z$, are associated following from Eq.~\rf{ab} that differ by sign only. Such combinations of wavenumbers are the most destabilizing in the sense that the magnetized Taylor-Couette flow is unstable at the highest possible Rossby number.

\begin{figure}[htp]
\begin{center}
\includegraphics[angle=0, width=0.35\textwidth]{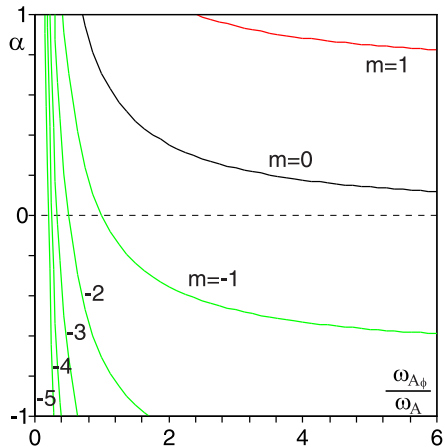}
\end{center}
\caption{Range of $\alpha$ for the different $m$ modes, in dependence on
$\beta/\alpha$: (red line) $m=1$, (black line) $m=0$, (green lines) $m=-1,-2,-3,-4,-5$. }
\label{fig3}
\end{figure}

In Fig.~\ref{fig3} we plot the positive branch of Eq.~\rf{ab} because the negative one is  simply its reflection about the horizontal coordinate axis.
We see now the HMRI mode ($m=0$) to start at $\omega_{A_{\phi}}/\omega_A=1/\sqrt{2}$ and
to remain for arbitrary large values of $\beta/\alpha$, although with an
ever decreasing wavenumber ratio, which would correspond to ever
increasing wavelengths in $z$ direction.
Again it is only the AMRI modes ($m=\pm1$) that,
for large $\beta/\alpha$, maintain a physically
sensible wavenumbers $\alpha=\pm1/\sqrt{2}$. The higher modes with $m\le-2$ are
obtained for smaller values of $\beta/\alpha$.

Hence, when the azimuthal magnetic field is directed along the basic flow that rotates counter-clockwise with respect to $z$-axis, among the modes that are MRI-unstable at the Liu limit there are only two (AMRI) that co-rotate with the flow $(m=1)$ and simultaneously propagate either along the positive or negative  $z$-direction. These modes are dominant when $\omega_{A_{\phi}}/\omega_A> 1+\sqrt{2}$, see the red curve in Fig.~\ref{fig3}. At moderate ratios $\omega_{A_{\phi}}/\omega_A$, the Liu limit is at the axisymmetric HMRI mode. When $\omega_{A_{\phi}}/\omega_A\rightarrow 0$, infinitely many modes with $m\le -1$ that propagate either in the negative or positive $z$-direction and counter-rotate with respect to the basic flow can cause instability at the Liu limit. Note, however, that at finite $\rm Re$ and $\rm Ha$ the highest modes will be inhibited, see Fig.~\ref{fig0}(b).

\section{Conclusion}

Using a short-wavelength approach, we have presented a unifying
picture of the
inductionless forms of MRI. We have identified a continuous function of
the maximum critical Rossby number that incorporates both
types of instability.
We were lead to the conclusion
that in the limit of small ratios of azimuthal to axial field there
should be inductionless MRI versions with higher $m$ modes, counter-rotating with respect to the basic flow, although this
needs further confirmation at least by a 1D linear stability analysis.
Most interestingly, the Liu limit has turned
out as being of quite universal significance, since the range of its
validity has been extended from the realm of axisymmetric HMRI  to that
of non-axisymmetric MRI versions.
Actually, soon after its derivation in the WKB
approximation, the relevance of
the Liu limit had been questioned by \cite{RH07} who
had found an apparent extension of this limit in global simulations
when at least one of the
radial boundary conditions was assumed to be electrically conducting.
Later, though, utilizing another definition of the notion ``quasi-Keplerian'' for
Taylor-Couette flows, the Liu limit was rehabilitated by \cite{P11}.
As a side remark, the determination of such limits is even more complicated by the
necessity to distinguish, for travelling waves as in HMRI, between
convective and absolute (or global) instabilities, which has been
thoroughly discussed by \cite{PG09} and which was
shown to be experimentally important by \cite{PRL}.

From the strictly astrophysical point of view, our support for the
Liu limit may appear disappointing, since it would exclude any relevance of the inductionless versions of MRI
to accretion disks with Keplerian rotation. A subtly question in this
respect is, however,
connected with the saturation mechanism of the MRI that could, possibly,
lead to modified shear profiles. With main focus on low $\rm Pm$ flows,
\cite{U} had asked for the possibility that the saturation of
MRI could lead to modified
flow structures within parts of steeper shear, sandwiched with parts
of shallower shear. By virtue of a possible sudden onset within such segments of steepening 
shear, the inductionless MRI versions could thus play a certain role
in real astrophysical settings.

\acknowledgments

This work was supported by the Deutsche Forschungsgemeinschaft
in frame of SFB 609 and SPP1488. O.N.K. gratefully acknowledges the support of the Japan Society for the Promotion of Science and of the Alexander von Humboldt Foundation.
We thank Rainer Hollerbach and G\"unther R\"udiger for numerous discussions on the various versions of MRI.


\begin{thebibliography}{99}

\bibitem[Balbus \& Hawley (1991)]{BH91} Balbus, S. A. \& Hawley, J. F.
1991, \apj, 376, 214


\bibitem[Balbus \& Henri (2008)]{BH08} Balbus, S. A. \& Henri, P. 2008,
\apj, 674, 408

\bibitem[Bilharz (1944)]{Bilharz44}
Bilharz, H. 1944, Z. angew. Math. Mech., 24, 77

\bibitem[Dobrokhotov \& Shafarevich (1992)]{DS92} Dobrokhotov, S. \& Shafarevich, A. 1992, Math. Notes, 51, 47

\bibitem[Eckhardt \& Yao (1995)]{EY95} Eckhardt, B. \& Yao, D. 1995, Chaos, Solitons \& Fractals, 5(11), 2073

\bibitem[Friedlander \& Lipton-Lifschitz (2003)]{FL03} Friedlander, S. \& Lipton-Lifschitz, A. 2003,
Localized instabilities in fluids. Handbook of Mathematical Fluid Dynamics, vol. II,
S.J. Friedlander and D. Serre, eds., Elsevier, 289

\bibitem[Fromang et al. (2007)]{FPLH} Fromang, S., Papaloizou, J.,
Lesur, G. \& Heinemann, T. 2007, A\&A,  476, 1123


\bibitem[Hattori \& Fukumoto (2003)]{HF2003} Hattori, Y. \& Fukumoto, Y. 2003, Phys. Fluids, 15, 3151

\bibitem[Hawley, Gammie and Balbus (1995)]{HGB95} Hawley, J. F., Gammie, C. F.,
Balbus, S. A., 1995, \apj, 440, 742


\bibitem[Hollerbach \& R\"udiger (2005)]{HR05} Hollerbach, R. \& R\"udiger,
G. 2005, Phys.\ Rev.\ Lett., 95, 124501

\bibitem[Hollerbach et al.\,(2010)]{HTR10} Hollerbach, R., Teeluck, V. \&
R\"udiger, G. 2010, Phys. Rev. Lett., 104, 044502


\bibitem[K\"apyl\"a \& Korpi (2011)]{KK} K\"apyl\"a, P. J. \& Korpi, M. J. 2011,
MNRAS, 413, 901


\bibitem[Kirillov \& Stefani (2010)]{KS10} Kirillov, O. N. \& Stefani, F. 2010, \apj,
712, 52

\bibitem[Kirillov \& Stefani (2011)]{KS11} Kirillov, O. N. \& Stefani, F. 2011,
Phys. Rev. E, 84, 036304

\bibitem[Krueger et al. (1966)]{KGD1966}
Krueger, E. R.,  Gross, A., Di Prima, R. C., J. Fluid Mech., 1966,  24(3), 521

\bibitem[Landman \& Saffman (1987)]{LS87} Landman, M. J. \& Saffman P. G.  1987, Phys. Fluids, 30, 2339

\bibitem[Lebowitz \& Zweibel (2004)]{LZ04} Lebovitz, N. R. \& Zweibel E. 2004, ApJ, 609, 301

\bibitem[Lesur \& Longaretti (2007)]{LL07} Lesur, G. \& Longaretti, P.-Y. 2007,
MNRAS, 378, 1471

\bibitem[Liu et al. (2006)]{LIU06} Liu, W., Goodman, J., Herron, I. \&
Ji, H. T. 2006, Phys.\ Rev.\ E, 74, 056302

\bibitem[Mizerski \& Bajer (2009)]{MB09} Mizerski, K. A. \& Bajer, K. 2009, J. Fluid. Mech., 632, 401

\bibitem[Nornberg et al. (2010)]{NO10} Nornberg, M. D., Ji, H., Schartman, E.,
Roach, E. \& Goodman, J. 2010, Phys. Rev. Lett., 104, 074501

\bibitem[Ogilvie \& Pringle (1996)]{OP} Ogilvie, G. I. \& Pringle, J. E. 1996,
MNRAS, 279, 151



\bibitem[Oishi \& Mac Low (2011)]{OML11} Oishi, J. S. \& Mac Low, M.-M. 2011,
\apj, 740, 18



\bibitem[Pessah \& Chan (2008)]{PH08} Pessah, M. E. \& Chan, C.
2008, \apj, 684, 498

\bibitem[Petitdemange et al. (2008)]{PDB08} Petitdemange, L., Dormy, E. \& Balbus, S. A.
2008, Geophys. Res. Lett., 35, 15305


\bibitem[Priede (2011)]{P11} Priede, J. 2011, Phys. Rev. E, 84, 066314

\bibitem[Priede \& Gerbeth (2009)]{PG09} Priede, J. \& Gerbeth, G. 2009,
Phys. Rev. E 79, 046310

\bibitem[Remillard \& McClintock (2006)]{RM06} Remillard, R. A. \& McClintock, J. E.
2006, Annu. Rev. Astron. Astrophys., 44, 49

\bibitem[R\"udiger and Hollerbach (2007)]{RH07} R\"udiger, G. \& Hollerbach, R.
2007, Phys. Rev. E, 76, 068301

\bibitem[Seilmayer et al. (2012)]{SE12} Seilmayer, M., Stefani, F., Gundrum, T.,
Weier, T., Gerbeth, G., Gellert, M. \&  R\"udiger, G. 2012, Phys. Rev. Lett.,
in press, arXiv 1112.2103


\bibitem[Sisan et al. (2004)]{SL04} Sisan, D. R., Mujica, N.,
Tillotson, W. A., Huang, Y. M., Dorland, A. B., Hassam, A. B.,
Antonsen, T. M.  \& Lathrop, D. P. 2004,
Phys. Rev. Lett., 93, 114502


\bibitem[Stefani et al. (2006)]{SGGRSSH} Stefani, F., Gundrum, T., Gerbeth,
G., R\"udiger, G., Schultz, M., Szklarski, J. \& Hollerbach, R.  2006,
Phys.\ Rev.\ Lett., 97, 184502

\bibitem[Stefani et al. (2007)]{SGGRSH} Stefani, F., Gundrum, T., Gerbeth,
G., R\"udiger, G., Szklarski, J. \& Hollerbach, R.  2007,
New J. Phys., 9, 295

\bibitem[Stefani et al. (2009)]{PRL} Stefani, F., G. Gerbeth, Gundrum, T., Hollerbach, R.,
Priede, J.,  R\"udiger, G., \& Szklarski, J.,  2009,
Phys. Rev. E., 80, 066303

\bibitem[Tayler (1973)]{TA73} Tayler, R. J. 1973, MNRAS, 161, 365

\bibitem[Terquem \& Papaloizou (1997)]{TQ97} Papaloizou, J. C. B. \& Terquem, C. 2007,
MNRAS, 287, 771

\bibitem[Turner \& Sano (2008)]{TS08} Turner, N. J. \& Sano, T. 2008, \apjl, 679, L131

\bibitem[Umurhan (2010)]{U} Umurhan, O. M. 2010, A\&A, 513, A47


\end{thebibliography}
\end{document}